# Topological based classification of paper domains using graph convolutional networks.

Idan Benami[1], Keren Cohen[1], Oved Nagar[1], Yoram Louzoun[1,2]

1- Department of Mathematics, Bar-Ilan University, Ramat Gan 52900, Israel
2- Correspondence and requests should be addressed to YL. louzouy@math.biu.ac.il

## Abstract

The main approaches for node classification in graphs are information propagation and the association of the class of the node with external information. State of the art methods merge these approaches through Graph Convolutional Networks. We here use the association of topological features of the nodes with their class to predict this class.

Moreover, combining topological information with information propagation improves classification accuracy on the standard CiteSeer and Cora paper classification task. Topological features and information propagation produce results almost as good as text-based classification, without no textual or content information. We propose to represent the topology and information propagation through a GCN with the neighboring training node classification as an input and the current node classification as output. Such a formalism outperforms state of the art methods.

The code is available at : https://github.com/louzounlab/graph-ml/tree/master/gcn

## Introduction and related work.

One of the central assumptions in node classification tasks is that neighboring nodes have similar classes. This has been extensively used in node classification tasks, and in machine learning based approaches to predict node colors. Such approaches are now often denoted by graph machine learning (i.e. machine learning where the input is a graph/network). Three main approaches have been proposed to take advantage of a graph in machine learning:

A) Regularize the output requiring that neighboring nodes should have similar classes and graph partitioning.
B) Use the graph to propagate labels and learn the best propagation or use the graph to project the nodes to real valued vectors and use those for supervised or unsupervised learning.
C) Use graph convolution network to learn the relation between the input of a node and its neighbors to its class.

### Output regularization

Methods based on the first approach (regularization of output or graph partitioning) include among others partitioning the graphs based on the eigenvalues of the Laplacian (assuming that nodes with the same partition have similar classes). The Laplacian of a graph is : $L = D - A$ , where $D$ is a diagonal matrix, with the sum of each row in the diagonal and $A$ the adjacency matrix. This Laplacian is often weighted by multiplying it by $D^{1/2}$ on the left and the right to normalize for degree (Dhillon, Guan et al. 2007) (Karypis and Kumar 1995). Other works have used variants of this idea, each using smoothness and graph distance differently (e.g., (Belkin and Niyogi 2004) (Sindhwani, Niyogi et al. 2005)). An alternative approach is to use quadratic penalty with fixed labels for seed nodes (Zhou, Bousquet et al. 2004) (Zhu, Ghahramani et al. 2003).

### Node Projection and propagation

For the second approach (diffusion of labels), multiple diffusion and information propagation models have been proposed (Rosenfeld and Globerson 2017);for example, DeepWalk (Perozzi, Al-Rfou et al. 2014), where a truncated random walk is performed on nodes. It then uses these sentences as an input to skipgram to compute a projection of each word into $\mathbb{R}^N$ maximizing the sentence probability. Planetoid (Yang, Cohen et al. 2016) also uses random walks combined with negative sampling. Duvenaud et al. used a translation of subgraphs to hash functions for a similar task in the context of molecule classifications

(Duvenaud, Maclaurin et al. 2015). A very similar approach was presented by Leskovech by projecting nodes minimizing the distance of neighbored nodes in a truncated random walk (Node2vec (Grover and Leskovec 2016)). The DNGR model (Cao, Lu et al. 2016) uses random walk to compute the mutual information between points (the PPMI-positive pointwise mutual information), and then a SVD decomposition to project into space. PPMI is a measure often used in information theory. PPMI was used for word representations in(Levy, Goldberg et al. 2015) and is a sparse high dimensional representation. Another possible approach is the projection of the input over the Laplacian eigenvectors, and the usage of the projection for classification, where either the graph itself is used (in such a case, the eigenvectors themselves are used) or an input to the graph was used. In such a case, a convolution with these eigenvectors was used (Masci, Boscaini et al. 2015, Monti, Boscaini et al. 2016). A Multi-Dimensional-Scaling (MDS) projection of the points in the graphs was also used for a similar goal (Belkin and Niyogi 2002, Levy, Goldberg et al. 2015). An alternative approach was inspired by word embedding methods (Mikolov, Sutskever et al. 2013) such as word2vec. These methods use the graph to define a "context" in relation to which the node embedding is constructed. When the data includes only the graph, the embeddings are used as features and fed into existing predictors (Perozzi, Al-Rfou et al. 2014). When the data includes node features, these embedding are used as an additional regularization term to a standard loss over the labeled nodes (Kipf and Welling 2016, Yang, Cohen et al. 2016). As such, these methods can be thought of as propagating *features* rather than labels. Refex (Henderson, Gallagher et al. 2011) defines local features to translate each node to a vector features and use those to predict classes.

### Graph convolutional network

Recently, Kipfs and collaborators, in a seminal work, propose a simplification of spectral based convolutions (Kipf and Welling 2016, Schlichtkrull, Kipf et al. 2017), and instead use a two-layer approach, which can be summarized as: softmax(A ReLU(AXW1)W2), where A is a normalized adjacency matrix $A = D^{-1/2}AD^{-1/2}$. They test their work on multiple graphs with labeled nodes including CiteSeer, Cora, Pubmed and Nell. Convolution approaches can also be used where the graph is used as a filter on the input. Most such convolutions are spectral (use the Laplacian eigenvectors). However, recent methods are based on random filters. Such convolutions include among other: *Atwood et al.* (Atwood and Towsley 2016), which define predetermined convolutions with powers of the adjacency matrix and then combine these powers using learned weights to maximize the classification precision of either the full graph or the classification of nodes. Bruna et al. *(Bruna, Zaremba et al. 2013)* provide a multi-level graph convolution with pooling, where at each stage nodes are merged into clusters using agglomerative clustering methods, and combine it with a pooling method to represent the different resolution of images. This has been extended (Henaff, Bruna et al. 2015, Bronstein, Bruna et al. 2016) to different convolutional kernels (mainly spectral, but also diffusion based kernels) and the classification of images, using ImageNet (see (Bronstein, Bruna et al. 2016) for a detailed review of all convolution methods). Vandergheynst and collaborators (see again review in (Bronstein, Bruna et al. 2016)) mainly use polynomial convolution in the spectral domain. Similar formalisms were used to study not only single snapshots, but also using recurrent networks time series of graphs, mainly again in image analysis (Seo, Defferrard et al. 2016).

### Proposed approach

A weakness of most of these approaches is the usage of the graph only as a similarity measure and ignoring any more complex features of topology of directed graphs, focusing on the above-mentioned assumption that proximity in the graph implies similarity in labels.

The main advances presented here are
  A) The directed graph topology-of each node contains information on its class and can be used to predict the class.
  B) When topology is combined with information propagation and convolution of external information, the combination outperforms all current methods.
  C) Even in the absence of external information, combining topology and propagation can produce an excellent classification.
  D) We propose an adaptation to the Kipf and Welling formailsm to show that topological features contain enough information to classify nodes. We then show that when combined with content, such methods outperform the state of the art methods, leading to a new method to classify nodes using GCN. This adaptation includes the introduction of the both symetric and anti-symetric components of the adjacency matrix in the learning step.

### Results

A typical example for node classification is the classification of the category of a manuscript from its content as characterized for example through its Bag Of Words (BOW) description. Two standards test sets used for this goal are the Cora and CiteSeer citation networks (Giles,

Bollacker et al. 1998). The current state of the art method to predict the category of a paper is through GCN, where the input for each node in the citation network is its Bag of Words (BOW), and the output of each layer is a function of the product of the adjacency matrix with the input and with weights optimized by the network:

(1) $X_{n+1} = \sigma(A \cdot X_n \cdot W_n)$

where $A$ is the adjacency matrix, $W_n$ are the weights from the $n$ layer, $X_n$ are the values of the nodes in this layer, and $\sigma(x)$ is a non-linear function. The input $X_o$ is a BOW and the output of the last layer is a soft-max used to determine the probability of each category given the input. The adjacency matrix is typically replaced by a normalized and symmetrized version (e.g., $D^{-\frac{1}{2}}(A + A^T + I)D^{-\frac{1}{2}}$, Other versions of GCN with a similar concepts have also been proposed (Schlichtkrull, Kipf et al. 2017) (Masci, Boscaini et al. 2015) (Henaff, Bruna et al. 2015) (Duvenaud, Maclaurin et al. 2015) (Kipf and Welling 2016).

While highly precise, such methods presume the existence of external information on the nodes, such as the BOW of each manuscript. In the absence of any such information, we propose that combining the relation between the node topology and its class with existing methods of class propagation can produce a similar classification accuracy.

To test that neighbor class and the node self-topology (as shall be further defined) are correlated with the node class, we produced two tests. We first computed the relative frequency of classes in neighbors, given the class of the node: $\frac{p(neighbor\ has\ class\ i | Current\ node\ has\ class\ j)}{P(node\ has\ class\ i)}$ (Fig 1C). In the absence of correlations, one would expect a flat value, while an absolute correlation would produce an identity matrix. In the Cora or Citeseer networks, the mass of the diagonal is 60 % of the mass (compared with an expected 15 % (Fig 1C)).

To test for the relation between node topology and class, we computed the average value of multiple topological features (Table 1) for nodes with a given class (in the current context manuscripts belonging to a certain field). One can clearly see (Fig 1B) that the average of many features varies according to the class of the node. When a Kruskal Wallis non-parametric test is performed to test for the relation between the node class (manuscript field) and the distribution of features, most features are associated with the node class (Fig 1A).

To test that topology and information propagation can be used to classify node classes, we introduced the topological features above, and the number of neighbors belonging to the training set with a given class as input to a Feed Forward Network. These two types of information by themselves can be used to classify the class of nodes quite precisely (Fig 2A).

In order to avoid having to explicitly compute topological features, which can be computationally expensive, a direct computation of the topology can be proposed. A simple way to describe such local features is though operations on products of the adjacency matrix. For example, the number of triangles $i\text{->}j$, $i\text{->}k$ and $j\text{->}k$, are the product of $a_{ij}$ and $b_{ik}; B = A * A^T$ (Fig 2B). Thus, instead of explicitly computing such features, one can use as input to the FFN combinations of these products on a one hot representation of the training set class. Formally, let us define for node $i$, the vector $v_i$, where $v_i^j$ is the number of neighbors of node $i$ that are in the training set and are of class $j$, and $V$ is the matrix of all vectors $v_i$. To this we add a last constant value to the vector, as shall be explained. We then use different combinations of $A * V, A^T * V, A * A^T * V$ .... as inputs to an FFN (see Methods).

When these products are applied to the last row (a constant value), they simply count sub-graphs. However, when multiplied by the other component, the sub-graphs composed of a specific class is counted (Fig 2B). The accuracy obtained for such products outperforms the only of explicit topological measures, or information propagation (Fig. 2A).

Such a formalism is easily amenable to a graph convolution network, by using $V$ as the input, but incorporating also the sign of second neighbors (i.e., $v_i^j$ would be the number of second neighbors of node $i$ that are in the training set and are of class $j$). Then instead of performing the product of $A$ with $V$, we use $V$ as the input a convolutional neural network. Note that such an input can then be combined with other inputs, such as the BOW, used by Kipf et al. Also., in order to incorporate the products of both $A$ and $A^T$, some reshaping must be performed within the network (see methods).

### Combination of BOW and topology outperforms BOW.

To test that adding such combinations improves the accuracy compared with state of the art methods, we have compared the accuracy of a network as proposed by Kipf et al. (Kipf and Welling 2016) (we have validated through an extensive grid search that the parameters proposed by Kipf et al. are indeed optimal for this architecture), with a similar network using as an input a concatenation of the BOW and the neighbor and second neighbor test category frequency. Indeed (Fig 3A), the combination of the topological propagation with the BOW improves the accuracy for all train sizes tested (except for the 15% and 25 % training fraction where no significant difference was observed). Note that the random accuracy in the Cora dataset is 20% and there is an inherent level of arbitrariness in the definition of scientific domains and thus a precision of 100 % cannot be reached.

While combining neighbor information with BOW outperforms state of the art methods, the neighbor information by itself produces results not far from that (less than 5% below) (Fig 3A). Moreover, for large training fractions, the purely topological outperforms the BOW based classification. Thus, even in the absence of any external information, a high precision can be obtained on node class classification, just by combining the network topology and the classification of neighboring nodes.

### The direction of edges is important

In contrast with images, directed networks have a complex topology where the direction of edges is of importance. Thus, explicitly incorporating the adjacency matrix and its transpose in the convolutional network is expected to produce better results than the symmetrized and self-edged adjacency matrix $\left(\frac{A+A^T}{2} + I\right)$. To test for that, we compared the performance of the two networks (each normalized as above). In all training set fractions, the full representation outperforms the symmetric representation or has equal accuracy (Fig 3A). Thus, it is not only the density of a class in the environment contributing to the probability of having the same class, but the explicit topology of the network. We have further tested whether explicitly adding the topological features to the input (instead of letting the network deduce the optimal network combination) would improve the accuracy. However, no difference was observed between the results with and without the explicit topological features.

### Discussion

Graph convolutional networks are typically viewed as methods to aggregate information from multiple distances (e.g., first neighbors, second neighbors, etc.). However, in contrast with images that are typically overlaid on a 2D lattice, graphs have a complex topology. This topology is highly informative of the properties of nodes and edges (Rosen and Louzoun 2016, Naaman, Cohen et al. 2018), and can thus be used to classify their classes.

In undirected graphs, the topology can often be captured by a distance maintaining projection into $\mathbb{R}^N$. Such a projection is often obtained by MDS methods (Kruskal 1964), or using supervised methods to minimize the distance between nodes with similar classes in the training set (Cao, Lu et al. 2016). In directed graphs, a more complex topology emerges from the asymmetry between incoming and outgoing edges (i.e., the distance between node *i* and node *j* differs from the distance between node *j* and node *i*), creating a distribution of subgraphs around each node often denoted sub-graph motifs (Milo, Shen-Orr et al. 2002). Such motifs have been reported to be associated with both single node/edge attributes as well as whole-graph attributes (Shen-Orr, Milo et al. 2002). We have here shown that in a manuscript assignment task, the topology around each node is indeed associated with the manuscript source.

In order to combine topological information with information propagation, we proposed a novel GCN where the fraction of second neighbors belonging to each class is used as an input, and the class of the node is compared to the softmax output of the node. This structure can be combined with external information about the nodes through a dual input structure, where the projection of the network is combined with the external information. We checked that such a formalism outperforms current state of the art methods in manuscript classification. Moreover, even in the absence of any information on the text, the network topology is enough to classify the manuscript almost as precisely as by reading the content of the manuscript.

The results presented here are a combination of information propagation and topology-based classification. While each of these two elements were previously reported, their combination into a single coherent GCN based classifier provides a novel content intendent method to classify nodes. With the current ever-increasing concerns about privacy, new content independent methods for node classification are required. The here proposed approach provides the currently optimal approach for such classification tasks.

## Methods

### Networks measures

Our goal is to use the graph structure to classify node colors. Hence, we compute features that are only based on the graph structure, ignoring any external content associated with each node. Those features are used to convert nodes into the appropriate network attribute vector (NAV) (Naaman, Cohen et al. 2018). Following is a list of attributes used. Note that other attributes may have been used with probably similar results.

- **Degree** -The first and easiest feature is the number of in & out (in case of directed graphs) edges.
- **Betweenness Centrality (Everett and Borgatti 1999).** Betweenness is a centrality measure of a vertex. It is defined by the numbers of shortest paths from all vertices that pass through the vertex. Formally: $g(v) = \sum_{s \neq v \neq t} \frac{\sigma_{st}(v)}{\sigma_{st}}$, where $\sigma_{st}$ is the total number of shortest paths from vertex '$s$' to vertex '$t$' and $\sigma_{st}(v)$ is the number of paths that pass through vertex '$v$'.
- **Closeness Centrality.** Closeness is a centrality measure of a vertex. It is defined as the average length of the shortest path between the vertex and all other vertices in the graph.
- **Distance distribution.** We compute the distribution of distances from each node to all other nodes using a Djekstra algorithm (Dijkstra 1959), and then use the first and second moments of this distribution.
- **Flow (Rosen and Louzoun 2014)** . we define the flow measure of a node as the ratio between the undirected and directed distances between the node and all other nodes. Formally:

$$F(u) = \frac{1}{|B(u)|} \sum_{v \in V} \frac{\overleftrightarrow{d(u,v)}}{\overrightarrow{d(u,v)}} \delta(u)$$

Where $\overrightarrow{d(u,v)}$ is the shortest path distance from $u$ to $v$ in a directed graph, and $\overleftrightarrow{d(u,v)}$ is the shortest path distance from $u$ to $v$ in undirected graph. $|B(u)|$ is the number of the vertices that can be reached in and out from vertex $u$ in any distance. The binary delta function deals with the case of small isolated group of vertices.

$$\delta(u) = \begin{cases} 1, & \frac{|B(u)|}{\max\{|B(u)|\}_{v \in V}} > threshold \\ 0, & otherwise \end{cases}$$

- **Attraction (Muchnik, Itzhack et al. 2007)** . Attraction Basin hierarchy is the comparison between the weighted fraction of the network that can be reached from each vertex with the weighted fraction of the network from which the vertex can be reached. Formally:

$$A(i) = \left( \sum_m \frac{N_{-m}(i)}{<N_{-m}>} \alpha^{-m} \right) \Big/ \left( \sum_m \frac{N_m(i)}{<N_m>} \alpha^{-m} \right)$$

where $N_{-m}(i)$ is the number of the vertices from which the vertex i can be reached in m directional edges and $<N_{-m}>$ is the average $N_{-m}(i)$ for all vertices i in the graph. The weight $\alpha^{-m}$ is introduced to balance the influence of the immediate and remote circles.

- **Motifs** Network motifs are small connected sub-graphs. We use an extension of the Itzchack algorithm (Itzhack, Mogilevski et al. 2007) to calculate motifs. For each node, we compute the frequency of each motif where this node participates.
- **K-cores (Batagelj and Zaversnik 2003).** A K-core is a maximal subgraph that contains vertices of degree *k* or more. Equivalently, it is the subgraph of *G* formed by repeatedly deleting all nodes of degree less than *k*.
- **Louvain community detection algorithm (Blondel, Guillaume et al. 2008).** The Louvain algorithm is a community detection algorithm. The algorithm works by optimization of modularity, which has scale value between -1 to 1. Formally, modularity is:

$$Q = \frac{1}{2m} \sum_{ij} \left[ A_{ij} - \frac{k_i k_j}{2m} \right] \delta(c_i, c_j)$$

where $A_{ij}$ represents the weight of the edge between *i* and *j*, $k_i = \sum_j A_{ij}$ is the sum of the weights of the edges attached to vertex *i*, $c_i$ is the community to which vertex *i* is assigned, the δ-function δ(*u*, *v*) is 1 if *u* = *v* and 0 otherwise and $m = \frac{1}{2}\sum_{ij} A_{ij}$.

### Datasets studied

We studied two citation networks – Cora & Citeseer (Giles, Bollacker et al. 1998). Citation graph's nodes are documents containing textual data, and the edges are citation links. The documents are split into several categories (classes) which we aim to predict. We analyzed two versions of these datasets which produced slightly different results. Either using the full graph which in both datasets is not connected or using the largest connected subgraph (see Table 2 for size of full and connected networks). In the main text we report the Cora results. The CiteSeer results are similar.

### Network structure

We compared the following two models of a GCN:

- **Multi-layer GCN**. We extended the GCN model developed by Thomas Kipf et al. (Kipf and Welling 2016). Each GCN layer is defined as $X_{n+1} = \sigma(A \cdot X_n \cdot W_n)$, where $A$ is the adjacency matrix, $X$ is the input from the previous layer, and $W$ are the weights of the current layer. The extension comes through the incorporation of an asymmetric adjacency matrix. In this model we do not lose the direction of the graph, which contains topological information. We incorporate the direction by taking the adjacency matrix (asymmetric in directed graph) and concatenate its transpose to it – creating a $2n \times n$ adjacency matrix. The dimension of the output of each layer is: $[(2N \times N) \cdot (N \times i_n) \cdot (i_n \times o_n)] = 2N \times o_n$., which in turn is passed to the next layer following a rearrangement of the output by splitting and concatenating it to change dimensions from $- 2N \times O_n$ to $N \times 2O_n$. The activation function for intermediate layers was the ReLu function, and the last layer was linear. Dropout rates of 40 %, and L2 regularization with a weight of 0.001 were used.
- Combined GCN: This model includes information from 3 sources: The adjacency matrix, the topological feature matrix, an external features matrix (in the Cora & Citeseer case, the bag-of-words features). As was the case above, the adjacency matrix is $\tilde{A} = A|A^T$, hence of dimensions: $2n \times n$. First, we pass the data matrix to a GCN-activation-dropout process, which leads to a $2n \times L_1$ output after the dropout. The two inputs (topology and external features) are then concatenated following a rearrangement of the processed data matrix by splitting in dimension 0 and concatenating in the dimension 1 – $2n \times L_1 \rightarrow n \times 2L_1$. Following the concatenation, an $n \times (2L_1 + T)$ matrix is obtained, which is passed forward to the Asymmetric GCN layers. The following layers are as above in the asymmetric GCN (see Fig 4 for comparison).

### Feed Forward Network

The results in Figure 2 are produced through a feed forward network with two internal layers of sizes 300 and 100 internal nodes and an output layer with the number of possible classifications (7 and 6 in CiteSeer and Cora, respectively). The nonlinearities were Relu's in the internal layers and a linear function in the output layer. An L2 regularization of 0.2 was used for all layers and a 10 % drop in our rate. The loss function was a categorical cross entropy as implemented in Keras with a TensorFlow backend.

### Code

The following links are the code parts to the graph-measures' calculations and the learning:

Topological GCN: [github.com/louzounlab/graph-ml/tree/master/gcn](github.com/louzounlab/graph-ml/tree/master/gcn)

Graph measure: [github.com/louzounlab/graph-measures](github.com/louzounlab/graph-measures)

### Statistical evaluation

The precision of the results was estimated through their accuracy, and the standard deviation of the accuracy was computed using 10 random training and test splits of the data. Methods were then compared using a non-parametric Mann-Whitney test.

# Figure Captions

Figure 1. A. Log p value of non-parametric Kruskal Wallis test for the association of each topological feature with the class of the manuscripts in CiteSeer. The *x* axis is the feature number (Table 1). One can clearly see that some features are highly associated with the manuscript class. B. Average of each topological feature for nodes belonging to a given class. All values were stacked and normalized to 1. An equal distribution would produce equally divided columns. C. Correlation of CiteSeer and Cora manuscript class with the neighboring manuscript class. The color in each row represents the fraction of nodes neighboring a given class that belong to all possible classes. One can clearly see the dominating diagonal representing the fact that neighboring nodes tend to have the same color. However, when topology is incorporated, an even more accurate estimate of the node color can be obtained.

Figure 2. A. Lower plot -- accuracy of the CiteSeer classification can be obtained when only topological features are used. Middle plot – accuracy when each node is classified using the distribution of classes in its neighbors. Upper plot – accuracy obtained when products of the neighbors and different combination of the adjacency matrix are used. B Example of sub-graph frequency through adjacency matrix products. Given the graph plotted on the left and the appropriate adjacency matrix (A), one can count the number of feed forward motifs ($x$->$y$ and $x$->$z$->$y$) originating from green and blue nodes through the product of A & A*A with the color one-hot matrix of the nodes (V). One can see that there is a total of 1 such triangle and it originates from a green node.

Figure 3. A. Comparison between the current state of the art (full dark line), our combined method (dashed dotted blue line), the same method when only topology is used (green dashed and red dotted lines). The dashed line is for the asymmetric network, while the red line is for the symmetric network. The *y* axis is the test accuracy and the *x* axis is the training set fraction. 3B. Network descriptions. Left side network GCN based on neighbor classification/topology. The network classification or topology is used as an input and the multiple GCN with either symmetric or full asymmetric networks are used. The nonlinearity in each internal layer is a Relu, and the output nonlinearity is a softmax when this is combined with external information as is the case for example for the BOW (right figure). This information is passed through a first GCN layer and the output of this layer is fed to the next layers (right network).

# Tables

| **All features** | **Neighbors** |
|---|---|
| attractor basin | first neighbors |
| average neighbor degree | second neighbors |
| betweenness centrality | |
| BFS moments | |
| closeness centrality | |
| eccentricity | |
| fiedler vector | |
| flow | |
| general | |
| K core | |
| load centrality | |
| louvain | |
| motif3 | |
| motif4 | |
| page rank | |

Table 1. List of features used. The first column depicts the list of features used for the topology. The second column is the list of features used when only neighbor information was used. Both types are described in detail in Methods.

| **Dataset** | **All graph** | | **Max connected subgraph** | | **Classes** |
|---|---|---|---|---|---|
| | **Nodes** | **Edges** | **Nodes** | **Edges** | |
| *"Cora"* | 2708 | 5429 | 2485 | 5209 | 7 |
| *"Citeseer"* | 3311 | 4703 | 2108 | 3745 | 6 |

Table 2. Node and edge number in networks studied in either the entire network, or only in the maximally connected component.

# Figures

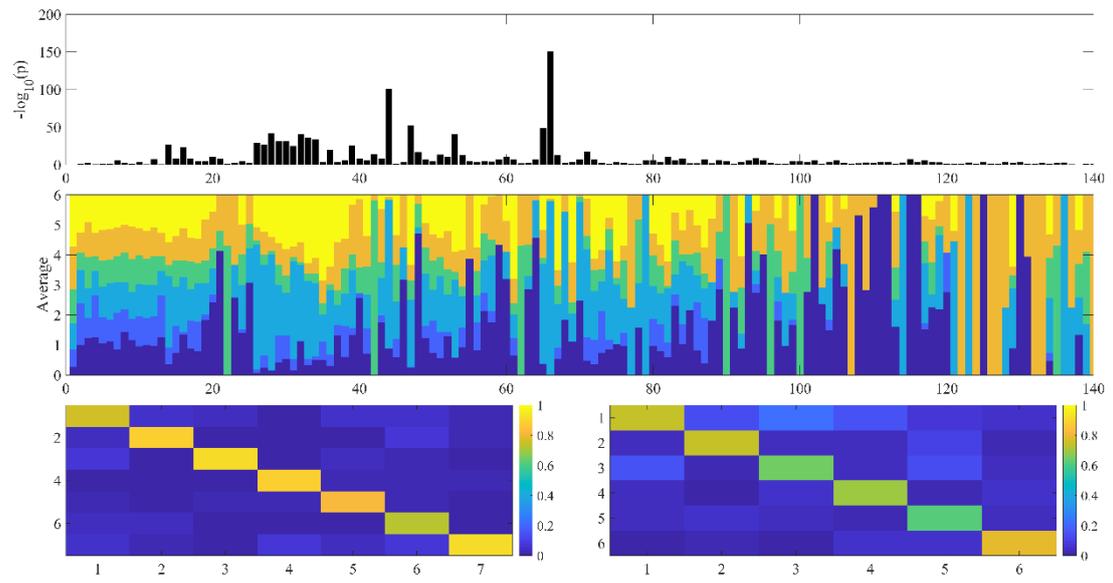

Figure 1.

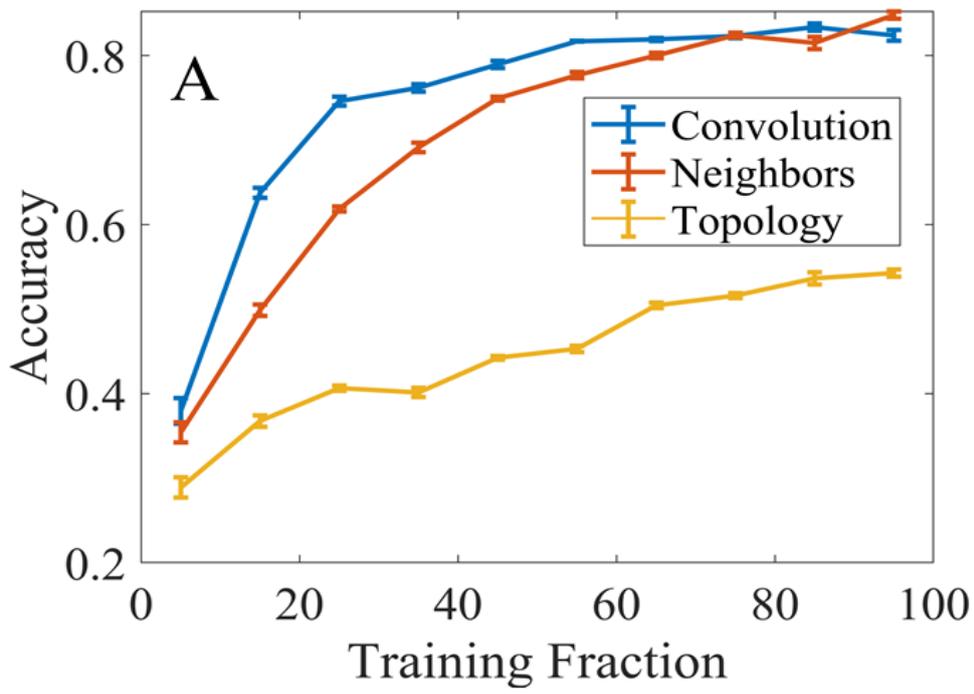
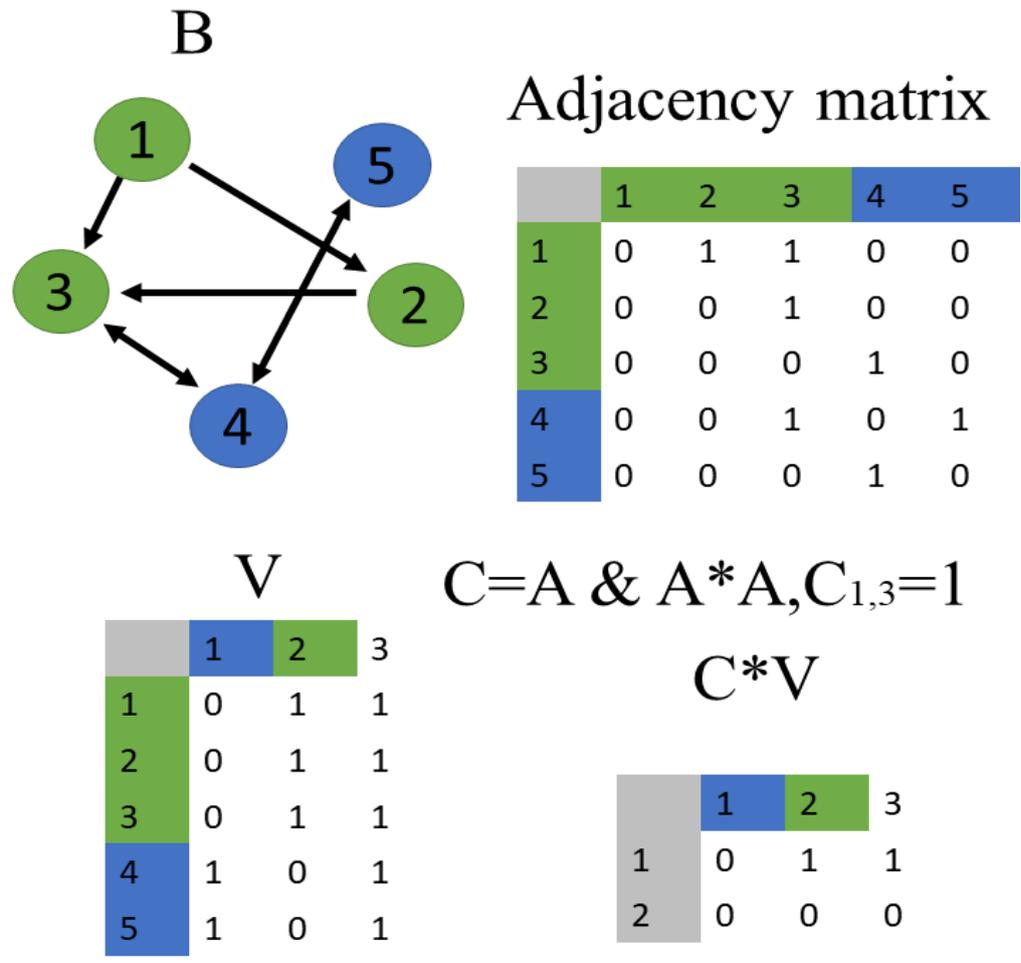

Figure 2

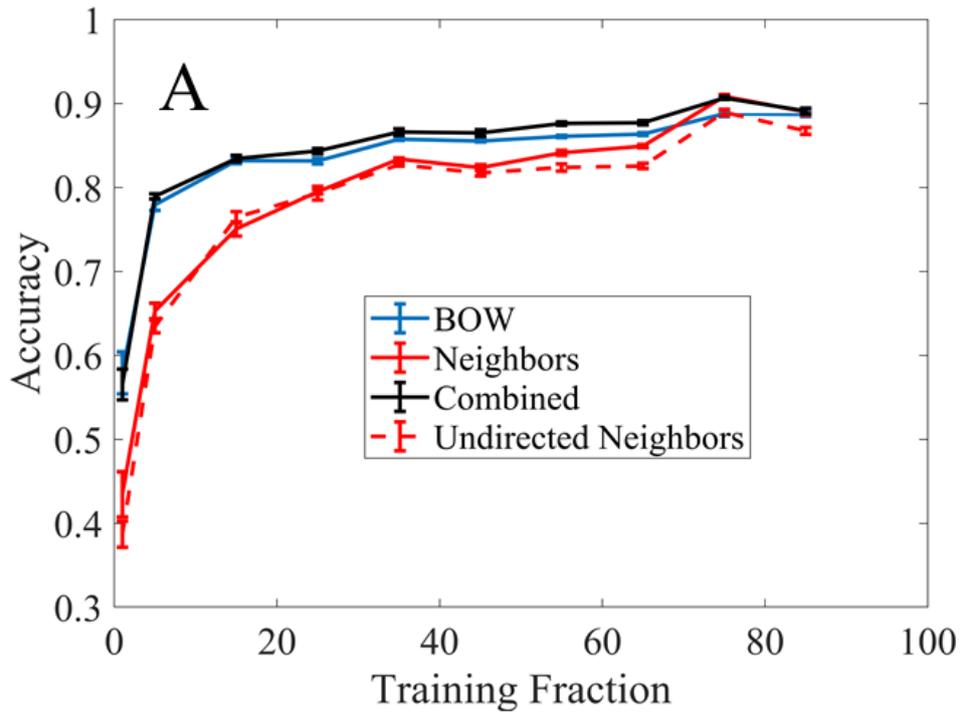

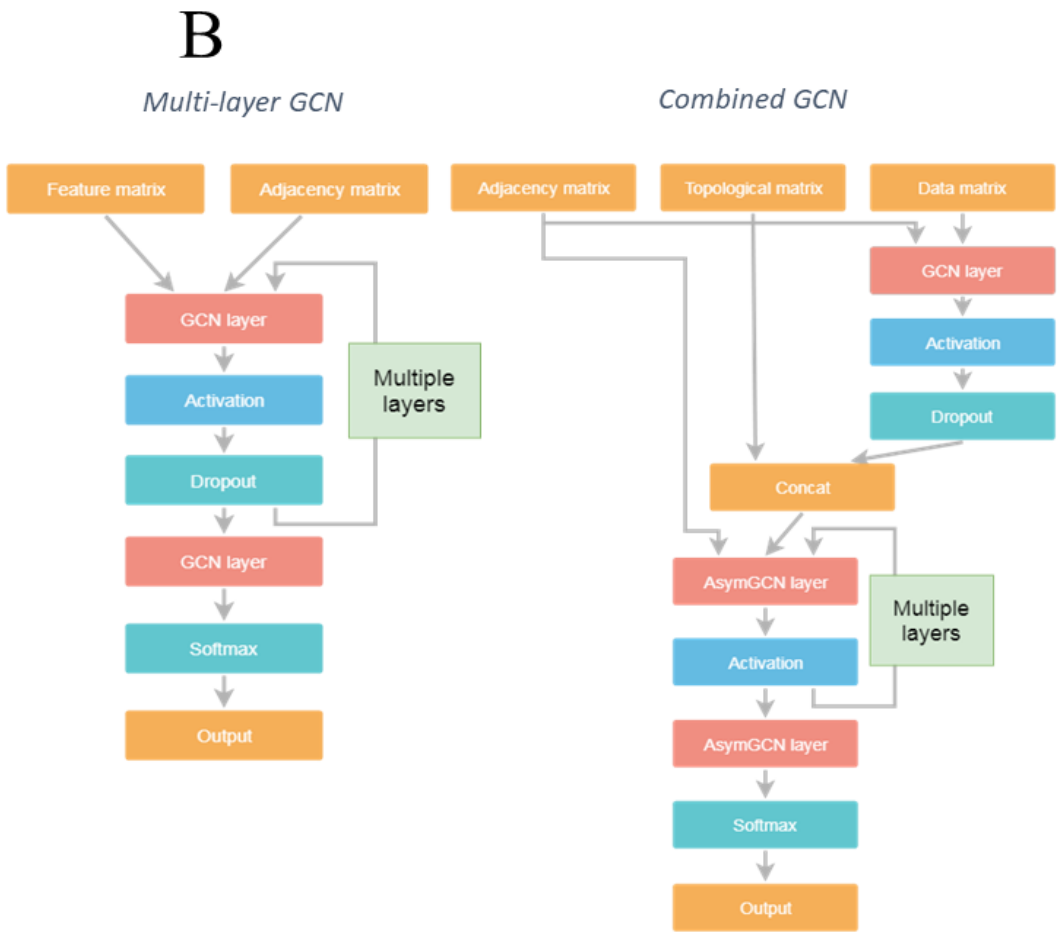

Figure 3